\documentclass[apj]{emulateapj}\slugcomment{Received 2013 October 9; accepted 2014 July 3; published 2014 August 14}
\usepackage{xcolor}
\usepackage{graphicx}
\usepackage{subfigure}
\usepackage{epsfig}
\usepackage{footmisc}
\begin{document}

\title{PHYSICAL NATURE OF THE [\ion{S}{2}]-BRIGHT SHELL NEBULAE N70 AND N185}

\author{Ning-Xiao Zhang\altaffilmark{1}}
\author{You-Hua Chu\altaffilmark{2}}
\author{R. M. Williams\altaffilmark{3}}
\author{Bing Jiang\altaffilmark{1,4}}
\author{Yang Chen\altaffilmark{1,4}}
\author{R. A. Gruendl\altaffilmark{2}}
\affil{$^1$School of Astronomy \& Space Science, Nanjing University, Nanjing 210093, China\\
$^2$University of Illinois at Urbana-Champaign, 1002 W. Green St., Urbana, IL 61801, USA\\
$^3$Columbus State University, 4225 University Avenue Columbus, GA 31907, USA\\
$^4$Key Laboratory of Modern Astronomy and Astrophysics, Nanjing University, Ministry of Education, China}

\begin{abstract}

N70 and N185 are two large ($\ge$100 pc in diameter) shell nebulae in the Large Magellanic Cloud (LMC).  Their high [\ion{S}{2}]/H$\alpha$ ratios rival those of supernova remnants (SNRs), but they are not confirmed as SNRs.  To study their physical nature, we have obtained \emph{XMM-Newton} X-ray observations and high-dispersion long-slit echelle spectroscopic observations of these two nebulae. The X-ray spectra of both nebulae can be well interpreted with an optically thin thermal ($\sim$0.2 keV) plasma with the average LMC abundance in a collisional ionization equilibrium.  N70 encompasses the OB association LH114.  Although N70 has a modest expansion velocity and essentially thermal radio emission, its diffuse X-ray luminosity ($\sim6.1\times10^{35}$ erg s$^{-1}$) is higher than that from a quiescent superbubble with N70's density, size, and expansion velocity; thus, N70 is most likely a superbubble that is recently energized by an interior SNR.  N185 does not contain any known OB association, and its X-ray luminosity is an order of magnitude lower than expected if it is a quiescent superbubble.  N185 has nonthermal radio emission and has high-velocity material expanding at nearly 200 km s$^{-1}$, similar to many known SNRs in the LMC. Its X-ray luminosity ($\sim1.9\times10^{35}$ erg s$^{-1}$) is also consistent with that of an evolved SNR.  We therefore suggest that N185 is energized by a recent supernova.
\end{abstract}

\keywords{ISM: bubbles -- ISM: lines and bands -- ISM: supernova remnants -- X-rays: ISM}

\section{Introduction}

Conventional optical surveys of supernova remnants (SNRs) use the criterion [\ion{S}{2}]/H$\alpha$ $\ge0.45$ \citep[e.g.,][]{1997ApJS..112...49M}, because the [\ion{S}{2}] $\lambda\lambda$6717, 6731 doublet emission is strong in the cooling zone behind SNR shocks.  In the Large Magellanic Cloud (LMC), a number of large shell nebulae were reported to have [\ion{S}{2}]/H$\alpha$ ratios $>$0.5 and even as high as 1.0 \citep{1977ApJ...212..390L,1999PASP..111..465S}, but their nature as SNRs could not be confirmed.  Two of these [\ion{S}{2}]-bright shell nebulae, LHA 120-N70 and LHA 120-N185 \citep[N70 and N185 for short;][]{1956ApJS....2..315H}, are particularly intriguing because of their extremely filamentary shell morphologies and high [\ion{S}{2}]/H$\alpha$ ratios in parts of their shell rims.  N70 is also known as DEM\,L301 and N185 as DEM\,L25 \citep{1976MmRAS..81...89D}.

The arguments against the SNR hypothesis for N70 include the following: (1) its $\ge$100 pc diameter is much larger than those of known SNRs; (2) its radio spectral index \citetext{$\sim -0.1$, \citealp{1981ApJ...250..103D}; $\sim -0.3$, \citealp{2002AJ....123..255O}; $-0.12\pm0.06$, \citealp{2014arXiv1404.3823D}} is more in accord with thermal than nonthermal emission; and (3) its expansion velocity \citetext{23 km\,s$^{-1}$, \citealp{1977ApJ...212..390L}; $\sim$70 km\,s$^{-1}$, \citealp{1981A&A....97..342R}} is much lower than those of SNRs. Furthermore, N70 encompasses the OB association LH114 \citep{1970AJ.....75..171L}, and thus N70 could be a superbubble blown by the fast stellar winds and supernovae (SNe) from LH114.

The strongest argument for the existence of SN explosion in N70 is the detection of diffuse X-ray emission by the \emph{Einstein X-ray Observatory}. As the diffuse X-ray luminosity of N70 is higher than that expected from a quiescent superbubble, it is suggested that N70 is a superbubble experiencing additional heating by recent SNe in the shell interior \citep{1990ApJ...365..510C}.  In an optical study, the H$\alpha$, [\ion{O}{3}], [\ion{N}{2}], and [\ion{S}{2}] images of N70 have been used to analyze the nebular morphology and excitation, and it is concluded that the stellar population in LH114 provides more ionizing radiation than needed and that the high [\ion{S}{2}]/H$\alpha$ ratios may result from low-velocity shocks \citep{1999PASP..111..465S}.

N185 is also large, 125$\times$100 pc.  It contains no cataloged OB associations \citep{1970AJ.....75..171L}, while one late-O and eight early-B stars are projected within its boundary \citep{1996ApJ...465..231O}.  Its expansion velocity, measured with Fabry--Perot interferograms, is $\sim$70 km\,s$^{-1}$ \citep{1982A&A...115...61R}, similar to that of N70 measured with the same instrument \citep{1981A&A....97..342R}.  The radio emission of N185, with a spectral index of $-0.8$, is apparently nonthermal \citep{2002AJ....123..255O}.  Both \emph{Einstein} and \emph{ROSAT} observations of N185 show a hint of diffuse X-ray emission that needs to be confirmed by deeper X-ray observations.  N185 has been suggested to be a SNR that is now photoionized by the blue stars within \citep{1982A&A...115...61R}.

To critically examine the physical nature of N70 and N185, we have acquired \emph{XMM-Newton} X-ray observations and high-dispersion long-slit spectroscopic observations of these two shell nebulae.  In this paper, we describe the observations in Section 2, analyze the physical properties of the hot gas in N70 and N185 in Sections 3 and 4, respectively, and discuss the origin of the hot gas in Section 5.

\begin{figure*}[tbh!]
\centering
\includegraphics[width=0.6\textwidth,angle=0,clip=]{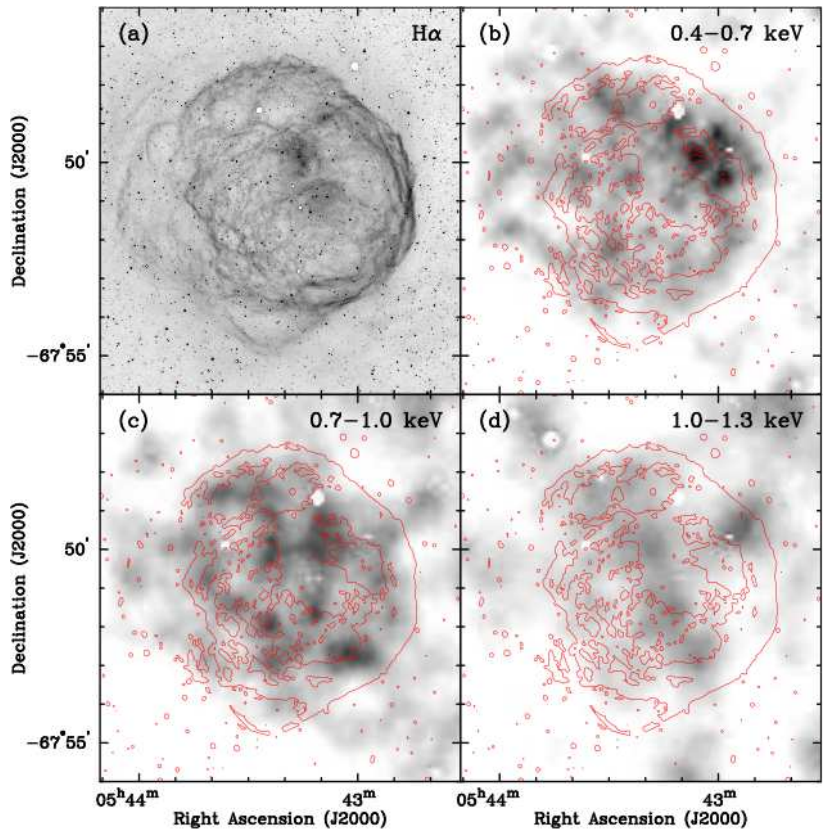}
\caption{Optical and X-ray images of N70: (a) H$\alpha$, (b) 0.4--0.7 keV, 
(c) 0.7--1.0 keV, and (d) 1.0--1.3 keV. H$\alpha$ contours are overplotted 
on the X-ray images to show their relative distributions. The H$\alpha$ image 
was obtained with the MOSAIC II camera on the CTIO 4 m telescope, and the X-ray 
images are from the \emph{XMM-Newton}.}
\label{figure1} 
\end{figure*}

\section{Observation and Reduction}

We have obtained \emph{XMM-Newton} X-ray observations and high-dispersion long-slit spectra of the H$\alpha$ and [\ion{N}{2}] lines of N70 and N185.  To complement the analysis of distribution and spectral properties of the diffuse X-ray emission from the shells, we have also used \ion{H}{1} observations and optical images.

\subsection{\emph{XMM-Newton} Observations}

Our \emph{XMM-Newton} observations of N70 were performed on 2008 January 26--27 (ObsID: 0503680201; PI: Williams) for 41.9 ks, and N185 on 2008 March 10--11 for 42.4 ks (ObsID: 0503680101; PI: Williams).  Both sets of observations were made in the full frame mode with the thin filter.  Three European Photon Imaging Cameras (EPIC), PN, MOS1, and MOS2, were used for the photon energy range of 0.15--15 keV.  Each camera has a field of view of $\sim$30$'$ and an on-axis angular resolution of $\sim$6$''$.

\begin{figure*}[tbh!]
\centering
\includegraphics[width=0.6\textwidth,angle=0,clip=]{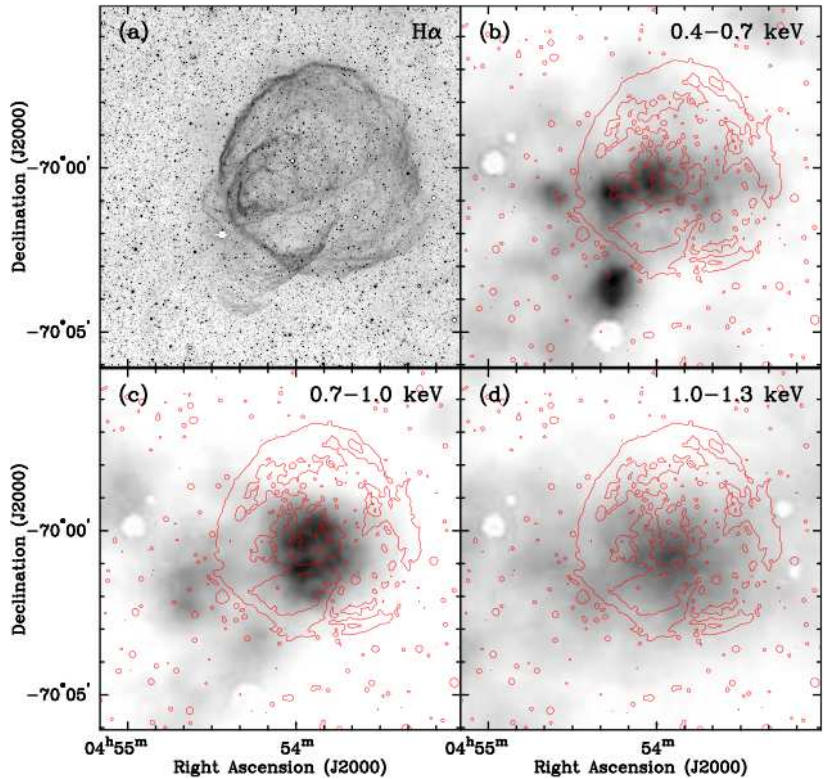}
\caption{Optical and X-ray images of N185: (a) H$\alpha$, (b) 0.4--0.7 keV, 
(c) 0.7--1.0 keV, and (d) 1.0--1.3 keV. H$\alpha$ contours are overplotted 
on the X-ray images to show their relative distributions. The H$\alpha$ image 
was obtained with the MOSAIC II camera on the CTIO 4 m telescope, and the X-ray 
images are from the \emph{XMM-Newton}.}
\label{figure2} 
\end{figure*}

The EPIC data were reprocessed with the Science Analysis System  software version 11.0.0 and the \emph{XMM-Newton} Extended Source Analysis Software (\emph{XMM}-ESAS) package \citep{2008A&A...478..615S,2008ApJ...674..209K}. After removing bad events and intervals of high background flares, the effective exposure times of the EPIC MOS1, MOS2, and PN are 23.7, 18.7, and 10.9~ks for N70, and 8.1, 8.3, and 4.8~ks for N185, respectively.  The quiescent particle background, soft proton contaminations, and exposure maps were all corrected using the \emph{XMM}-ESAS package.  Finally, point sources were detected and excised, and the resultant maps of the diffuse X-ray emission were adaptively smoothed.  The MOS1, MOS2, and PN combined images in the 0.4--0.7, 0.7--1.0, and 1.0--1.3 keV bands are shown in Figure \ref{figure1} for N70, and in Figure \ref{figure2} for N185.

\begin{figure*}[tbh!]
\centering
\includegraphics[width=0.8\textwidth,angle=0,clip=]{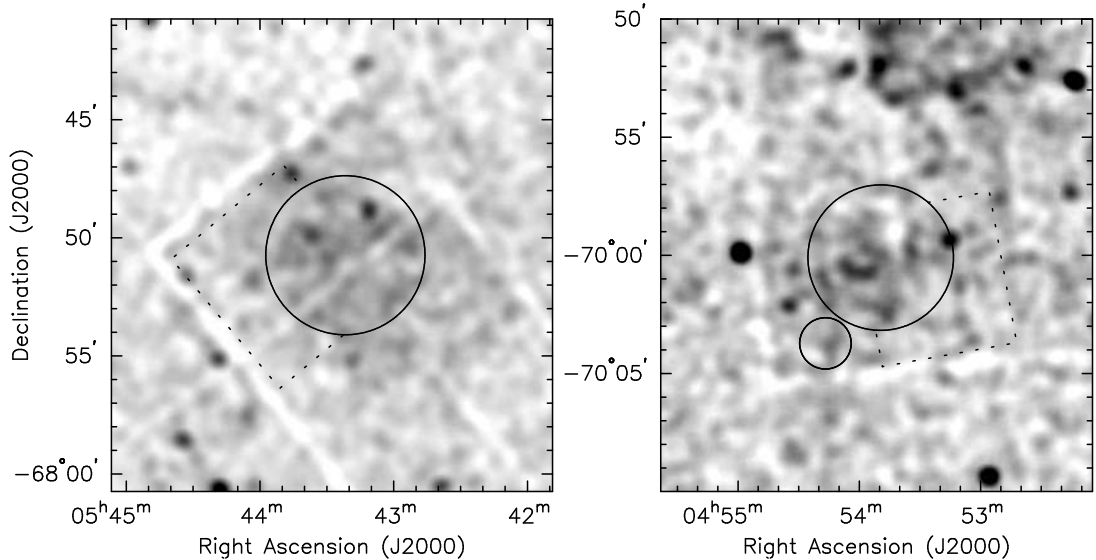}
\caption{X-ray images of N70 (left panel) and N185 (right panel) with source (circle) and background (box) regions overplotted. Two circular regions surrounding the shell and blowout, respectively, are defined in N185. These two regions together are used as the source region in the extraction of spectra, but the hardness ratios of these two regions are determined separately and compared.}
\label{figure3} 
\end{figure*}

To extract spectra from the EPIC observations, we define the source and background regions as shown in Figure \ref{figure3}.  Because of the small number of counts detected, a single source region is selected to include all diffuse emission in each object: the source region of N70 encompasses its entire shell, and the source region of N185 encompasses both the entire shell and the blowout.\footnote{The source region of N185 consists of two circular regions that surround the shell and the blowout, respectively. The X-ray spectra of N185 are extracted from these two circular regions combined, but the hardness ratios of these two circular regions are determined separately and compared in Section 4.2.}  In order to minimize the instrumental background, the background region is selected to be on the same CCD as the source region.  The total background-subtracted counts in MOS1, MOS2, and PN are 1141, 1182, and 2731 for N70, and 295, 240, and 272 for N185, respectively.  In the case of N185, the numbers of counts are small, thus the MOS1 and MOS2 data are combined to extract a spectrum. The background-subtracted spectra are then adaptively binned to achieve a signal-to-noise ratio (S/N) of at least 3.  The analysis and model fits to these spectra are described later in Sections 3.2 and 4.2.

\subsection{High-dispersion Echelle Observations}

Both N70 and N185 were observed with the echelle spectrograph on the Blanco 4\,m telescope at Cerro Tololo Inter-American Observatory (CTIO).  The observations were made in a long-slit, single-order mode, in which a broad H$\alpha$ filter was used and a flat mirror replaced the cross dispersor, resulting in a coverage of the H$\alpha$ and [\ion{N}{2}] $\lambda\lambda$6548, 6584 lines over a slit length of $\sim4'$.  A 79 lines mm$^{-1}$ echelle grating was used.

\begin{figure*}[tbh!]
\centering
\includegraphics[width=0.7\textwidth,angle=0,clip=]{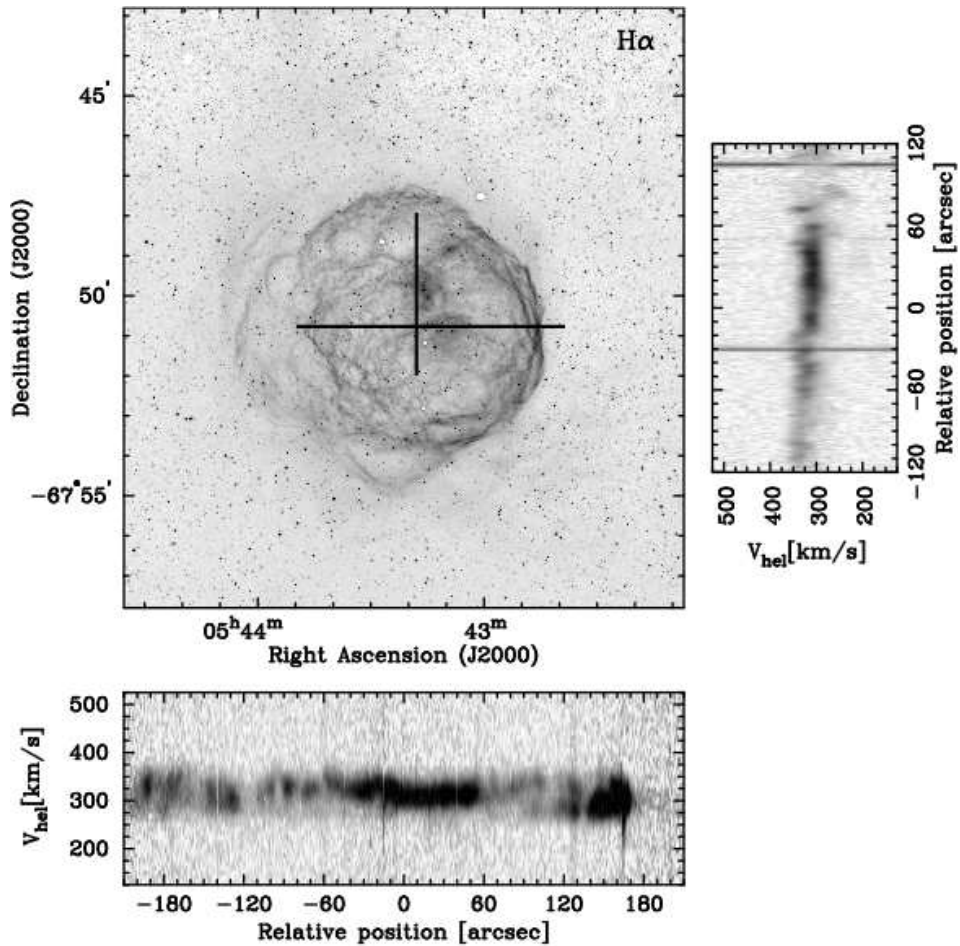}
\caption{H$\alpha$ image and H$\alpha$ echellograms of N70. The orientations of 
the echellograms are parallel to the slit positions marked on the H$\alpha$ image, 
but the image scales are different. The dispersion axis is labeled with 
heliocentric velocities.}
\label{figure4} 
\end{figure*}

N70 was observed on 1986 November 22 and 1988 January 11.  The Air Schmidt camera and a GEC 385$\times$576 CCD were used to record data; the 22 $\mu$m pixel size corresponded to 0\farcs635 along the slit and $\sim$0.21 \AA\ along the dispersion.  A slit width of 1\farcs65 was used and the resultant instrumental FWHM was 21$\pm$1 km\,s$^{-1}$ at the H$\alpha$ line.  Two consecutive east--west slit positions were observed in 1986 \citep{1988AJ.....95.1111C} and a north--south slit position was observed in 1988.  The long-slit echelle spectra are shown in Figure \ref{figure4} with slit positions marked on an H$\alpha$ image of N70.

\begin{figure}[tbh!]
\centering
\includegraphics[width=0.45\textwidth,angle=0,clip=]{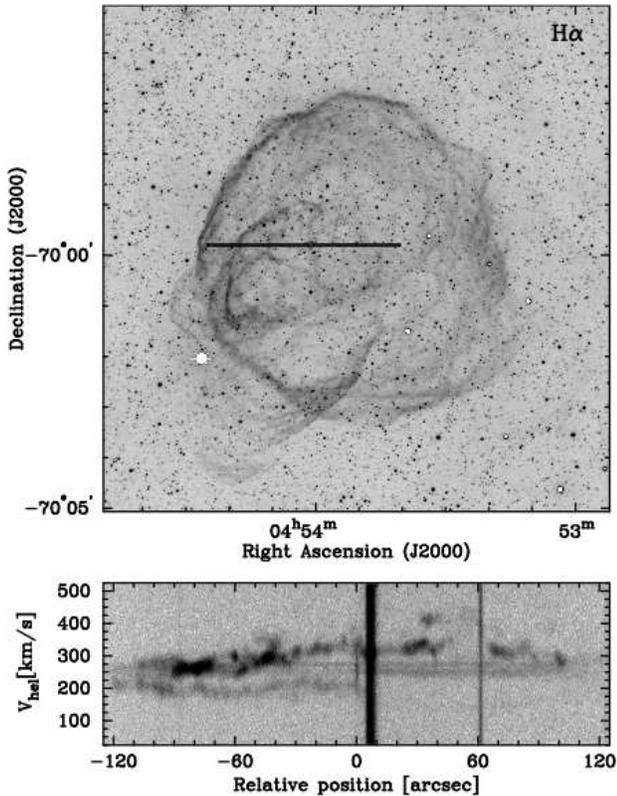}
\caption{H$\alpha$ image and H$\alpha$ echellograms of N185. The orientations of 
the echellogram is parallel to the slit position marked on the H$\alpha$ image, 
but the image scales are different. The dispersion axis is labeled with 
heliocentric velocities.}
\label{figure5} 
\end{figure}

N185 was observed on 1995 January 20.  The red long focus camera and a Tek 2048 $\times$ 2048 CCD were used to record data; the 24 $\mu$m pixel size corresponded to 0\farcs267 along the slit and 0.082 \AA\ along the dispersion.  A 1\farcs65 slit width was used and the instrumental FWHM was $\sim$14 km\,s$^{-1}$ at the H$\alpha$ line.  Only one east--west slit position was observed.  The long-slit echelle image and the slit position on an H$\alpha$ image of N185 are shown in Figure \ref{figure5}.

The echellograms in Figures 4 and 5 show very different velocity structures between N70 and N185.  N70 shows mostly unsplit line with centroid velocity variations within the range of $V_{\rm hel}$ = 280--340 km s$^{-1}$; only the western part of N70 (relative position 60$''$--150$''$ along the east--west slit) show a velocity split of $\sim$70 km s$^{-1}$.  From these velocity structures, we can conclude that the shell of N70 cannot be expanding faster than 35 km s$^{-1}$.  (Note that the center of expansion is not the geometric center of N70; therefore, no geometric correction is needed for the local expansion velocity.)  In contrast, N185 shows velocity splits throughout the slit length, consistent with a global expansion.  The receding side of N185 is detected at velocities from $V_{\rm hel}$ = 260 to 320 km s$^{-1}$ with additional high-velocity features reaching $V_{\rm hel} \sim$ 420 km s$^{-1}$, while the approaching side is detected at $V_{\rm hel}$ = 220--180 km s$^{-1}$.  In addition to these approaching and receding components of the N185 shell, there is a stationary component at $V_{\rm hel}$ $\sim$ 250 km s$^{-1}$, which can either be an irrelevant line-of-sight component or associated with the ambient medium of N185.  The systemic velocity of N185 is most likely in the range of $V_{\rm hel}$ = 240--250 km s$^{-1}$, considering the centroid velocity at the shell rim and the stationary component.  The bulk expansion velocity of N185 is thus at least $\sim$70 km s$^{-1}$ and expansion velocities up to $\sim$200 km s$^{-1}$ are present in certain parts of the nebula.  Therefore, we will adopt an expansion velocity of $\le$35 km s$^{-1}$ for N70 and $\ge$70 km s$^{-1}$ for N185.

The velocity structures of N70 and N185 determined from our echelle data are at odds with previous studies using Fabry--Perot interferograms, which reported expansion velocities of 70 km s$^{-1}$ for both N70 and N185 \citep{1981A&A....97..342R,1982A&A...115...61R}.  These differences can be explained by the deficiencies of Fabry--Perot interferograms: (1) the velocity information in Fabry--Perot interferograms can be erroneously modulated by the nebula's surface brightness, and this is particularly problematic for narrow filaments that do not fill the entire width of an interference fringe; (2) the limited free spectral range (FSR) causes aliasing among velocity components whose velocity differences are multiples of the FSR; and (3) the interferograms were recorded through an image tube onto photographic plates, which are not as sensitive as CCDs.  In the case of N70, the interferograms show that line splits are only seen in localized regions (Figure~1 of Rosado et al.\ 1981), consistent with our echelle observations; however, their expansion velocity of 70 km s$^{-1}$ was based on an optimistic fit of a velocity-radius ellipse to data points that clearly do not represent a hollow expanding shell (Figure\ 3 of Rosado et al.\ 1981).  In the case of N185, the 100 km s$^{-1}$ FSR of the interferogram precludes the detection of high-velocity features that are clearly detected in our long-slit echelle observations; therefore, the highest possible expansion velocity reported by \citet{1982A&A...115...61R} corresponds to our lower limit of N185's expansion velocity.

\subsection{Complementary Data}



Optical images of N70 and N185 are available from the Magellanic Cloud Emission Line Survey (MCELS; Smith and MCELS Team 1999).  The [\ion{O}{3}] images are useful in showing high-excitation regions and the [\ion{S}{2}] images reveal ionization fronts as well as cooling regions behind shocks.  The MCELS images have an angular resolution of $\sim$2$''$.  For a higher angular resolution ($\le$1\farcs0), we use our H$\alpha$ images of N70 and N185 taken with the MOSAIC II camera on the CTIO Blanco 4\,m telescope.

To examine the distribution of \ion{H}{1} in the vicinity of N70 and N185, we use the 21 cm line data cube of the LMC made with combined Australia Telescope Compact Array observations and Parkes single-dish observations \citep{2003ApJS..148..473K}.  Both the \ion{H}{1} column density map and iso-velocity maps are compared with the surface brightness variations of the diffuse X-ray emission in and around N70 and N185 to assess the distributions of foreground absorption (see Sections 3.1 and 4.1).

\section{Hot Gas in N70}

\subsection{Spatial Distribution}

N70 is a faint diffuse X-ray source.  To examine the spatial distribution of its X-ray emission, we extract images in the 0.4--0.7, 0.7--1.0, and 1.0--1.3~keV bands, avoiding 1.4--1.8 keV due to the instrumental fluorescence lines (Al K$\alpha$ at 1.49~keV and Si K$\alpha$ at 1.77~keV).  In addition, the image above 1.8 keV shows no appreciable emission, so we ignore the energy band above 1.8 keV.  The three X-ray images are displayed along with an H$\alpha$ image of N70 in Figure \ref{figure1}.  The contours of H$\alpha$ emission are overplotted on the X-ray images for comparison of their relative distributions.  In the 0.4--0.7 and 0.7--1.0 keV bands, it can clearly be seen that elevated diffuse X-ray emission exists within the H$\alpha$ shell.  In the 1.0--1.3 keV band, there is an apparent strip of diffuse emission within N70, but there are also patches of emission exterior to N70.  It is not clear whether the 1.0--1.3 keV emission projected within N70 is exclusively associated with N70; thus, we will focus on the 0.4--0.7 and 0.7--1.0 keV images for spatial analysis.  These two soft energy bands are called S1 and S2 in this paper.

The diffuse X-ray emission in the S1 and S2 bands, at a level of 70$\sigma$ over the background, is well confined within the H$\alpha$ shell of N70.  The bright diffuse X-ray emission is associated with the bright parts of the H$\alpha$ shell, and the faint diffuse X-ray emission extends toward the faint H$\alpha$ arcs to the east.  The S1 band image shows three additional patches of diffuse X-ray emission on the east exterior to the H$\alpha$ contours, but a closer examination of the H$\alpha$ image shows that these diffuse X-ray patches are still enclosed within H$\alpha$ filaments that are fainter than the lowest contour plotted.

\begin{figure*}
\centering
\includegraphics[width=\textwidth,angle=0,clip=]{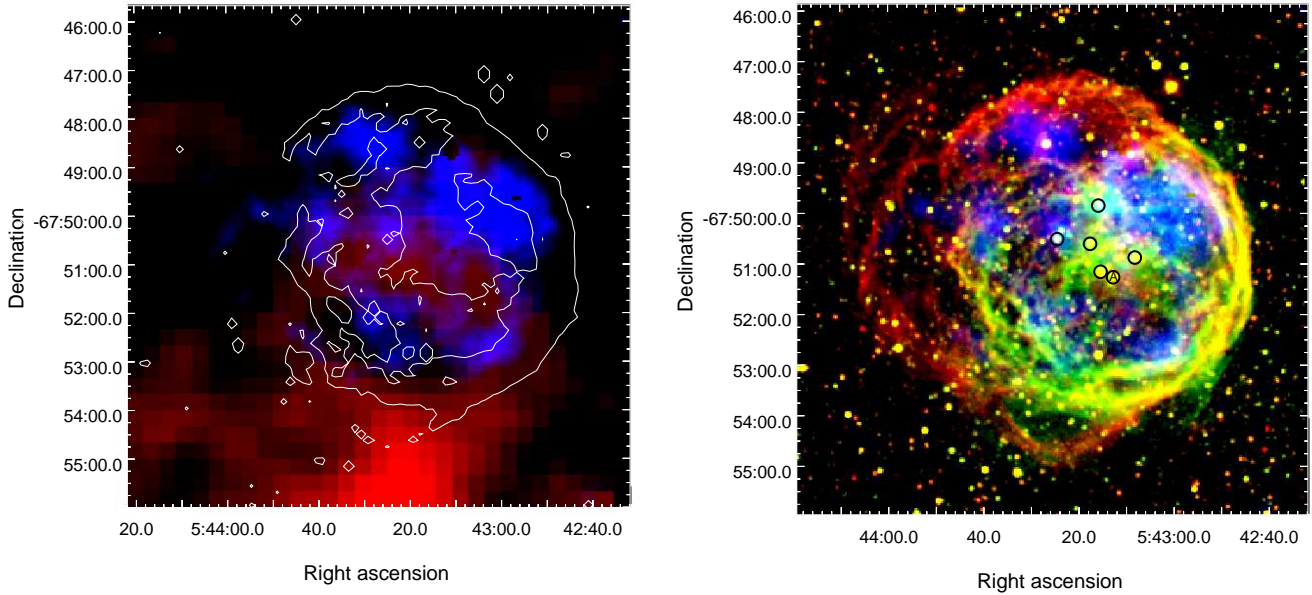}
\caption{Left panel: Color composite of \ion{H}{1} column density map (red), 0.4 -- 1.0 keV 
X-ray image (blue), and H$\alpha$ contours of N70. Right panel: Color composite of [\ion{S}{2}] (red), [\ion{O}{3}] (green), 
and 0.4 -- 1.0 keV X-ray (blue) images of N70. The six known massive stars are 
marked with circles and the earliest O star is marked with an "A" in the 
circle.}
\label{figure6} 
\end{figure*}

The diffuse X-ray surface brightness is not uniform within N70.  The background emission in the areas surrounding N70 shows variations, too.  To investigate whether the variations in X-ray surface brightness are caused by uneven foreground absorption, we compare the \ion{H}{1} column density map with the X-ray image in the 0.4--1.0 keV band in the left panel of Figure \ref{figure6}.  It is striking that an east--west elongated \ion{H}{1} cloud is coincident with the region of X-ray minimum, indicating that foreground absorption may indeed play a significant role in the nonuniform surface brightness of the diffuse X-rays in N70.

We have examined the \ion{H}{1} channel maps and find that \ion{H}{1} is detected at the position of N70 over $V_{\rm hel}$ = 295--305 km\,s$^{-1}$.  The ionized gas in N70 shows similar velocities (see Figure\ 4); therefore, N70 may well be associated with the same \ion{H}{1} complex.  The lack of organized expansion structure in the \ion{H}{1} around N70 is probably caused by N70's being density-bounded with its surrounding gas completely ionized \citep{2002AJ....123..255O}.  Indeed, our deep H$\alpha$ image of N70 shows diffuse H$\alpha$ emission exterior to the shell structure, indicating an extended ionized gas region.  The \ion{H}{1} velocity field may be further complicated by its location on the western rim of the \ion{H}{1} supergiant shell 23 \citep{2003ApJS..148..473K}.

\begin{figure}
\includegraphics[width=0.3\textwidth,angle=270,clip=]{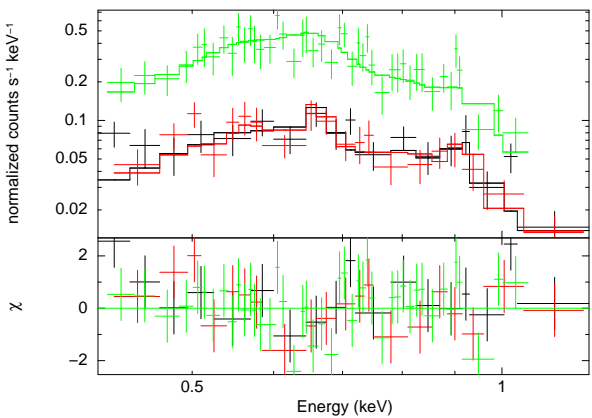}
\caption{\emph{XMM-Newton} EPIC X-ray spectra of N70. The MOS1, MOS2, and PN 
spectra are plotted in black, red, and green, respectively. The best-fit model 
is overplotted. See the text for more details.}
\label{figure7} 
\end{figure}

To compare the distribution of diffuse X-ray emission with that of massive stars in N70, the right panel of Figure \ref{figure6} shows a color composite of N70, with [\ion{O}{3}] in green, [\ion{S}{2}] in red, and 0.4--1.0 keV X-ray in blue. The six known O stars in N70 are marked. It can be seen that the [\ion{O}{3}] image exhibits patchy emission near the massive stars and a filamentary shell structure centered on an O3If star (R.A.=$5^{h}43^{m}13^{s}$, decl.=67$^{\circ}$51$'$16$''$), currently the most massive star in LH114. The centering of enhanced [\ion{O}{3}] emission on massive stars indicates photoionization by stellar UV fluxes.  The [\ion{S}{2}] emission, on the other hand, shows the overall shell structure of N70 and encloses the diffuse X-ray emission, suggestive of a common origin for the [\ion{S}{2}] emission and the X-ray-emitting hot plasma.

\subsection{X-Ray Spectral Analysis}

\begin{center}
\begin{deluxetable}{lcc}
\tabletypesize{\footnotesize}
\tablecaption{Spectral Fitting Results for N70 and N185 with $90\%$ Confidence ranges. }
\tablewidth{0pt}
\tablehead{
\colhead{Model} &\colhead{N70} &\colhead{N185}
}
\startdata
phabs ($N_{\rm H}$) ($10^{22}$) & $0.31_{-0.07}^{+0.06}$ & $0.22$\\
kT (keV) & $0.19_{-0.01}^{+0.01}$ & $0.20_{-0.01}^{+0.02}$\\
norm ($10^{-3}$) & $4.82_{-2.10}^{+3.33}$ & $1.42_{-0.30}^{+0.30}$\\
$\chi_{\nu}^{2}/$dof & 0.96 (77.01/80) & 1.36 (31.24/23)\\
Flux ($10^{-12}$erg cm$^{-2}$ s$^{-1}$)$^{a}$& $2.17_{-0.95}^{+1.50}$ & $0.68_{-0.14}^{+0.14}$\\
$L_{\rm X}$ $10^{35}$ erg\,s$^{-1}$ & $6.12_{-2.68}^{+4.23}$ & $1.92_{-0.40}^{+0.40}$
\enddata
\tablenotetext{a}{\phantom{0} The unabsorbed fluxes are in the 0.4--$1.5$ keV band.}
\end{deluxetable}
\end{center}

To assess the physical properties of the hot gas in N70, we model the background-subtracted adaptively binned X-ray spectrum of the diffuse emission within N70.  Figure \ref{figure7} shows the spectra individually extracted from the MOS1, MOS2, and PN.  While the PN observation detected more counts, the MOS spectra show better the line emission of \ion{O}{7} at 0.57 keV, \ion{O}{8} at 0.65 keV, and \ion{Ne}{9} near 0.92 keV.  The presence of the lines justifies the use of thermal plasma emission models.  Thus, we use the absorbed apec model and the XSPEC package (version 12.7.1) to fit the X-ray spectrum of N70 for the energy range of 0.4--1.3 keV.  The energy cutoff at 1.3 keV is selected because there is no appreciable emission above this energy.  We fix the abundance at 1/3 solar, which is appropriate for the LMC \citep{1992ApJ...384..508R}.  The MOS1, MOS2, and PN spectra are fit jointly, and the results are presented in Table 1.

The absorption column density from the best fit is $3.1 \pm 0.6 \times10^{21}$ H-atom cm$^{-2}$.  This can be compared with that implied by the extinction toward the OB stars in N70.  Most of the OB stars studied by \citet{1996ApJ...465..231O} have $E(B-V)$ in the range of 0.1--0.2 mag, corresponding to $A_V$ of 0.32--0.64 mag.  The Milky Way contribution to this extinction is $A_V^{\rm MW} \sim 0.206$ mag \citep{2011ApJ...737..103S}, and the LMC contribution is $A_V^{\rm LMC} \sim$ 0.114--0.434 mag.  Using the gas to dust ratio of $N_{\rm H}/E(B-V)$ = $5.8\times10^{21}$ H-atom cm$^{-2}$ mag$^{-1}$ for the Milky Way \citep{1978ApJ...224..132B} and $2.4\times10^{22}$ H-atom cm$^{-2}$ mag$^{-1}$ for the LMC \citep{1986AJ.....92.1068F}, the above extinction implies a H column density of 1.2--3.6$\times$10$^{21}$ H-atom cm$^{-2}$ toward most OB stars in N70. This column density is in good agreement with the best-fit X-ray absorption column density, considering the non-uniform coverage of \ion{H}{1} over N70.

The density of the hot gas in N70 can be derived from the normalization factor ($A$) of the best-fit model of the X-ray spectra:$$A = \frac{10^{-14}}{4\pi D^2} \times n_{\rm e} n_{\rm H} f V~ {\rm cm}^{-5} $$ where $D$ is the distance to N70, $n_{\rm e}$ and $n_{\rm H}$ are electron and hydrogen number densities, $f$ is the filling factor, and $V$ is the volume of the N70 superbubble, all in cgs units.  Adopting an LMC distance of 50 kpc and a shell radius of 50 pc for N70 and assuming a He/H number ratio of 10\%, the hydrogen atom density is $n_{\rm H}$ = 0.1 $f^{-1/2}$ cm$^{-3}$.  The total mass of the X-ray emitting gas is $1.1 \times 10^{3}$ $f^{1/2}$ $M_\odot$.  Such a large amount of mass suggests that the hot gas is mostly shock-heated interstellar matter.

\section{Hot Gas in N185}

The \emph{XMM-Newton} observations of N185 were hampered by high background flares, and the useful exposure time was significantly shortened.  Because of its shorter exposure and lower surface brightness, diffuse X-ray emission from N185 is detected at a much lower S/N than that from N70.  The spatial and spectral analyses of N185 are quantitatively much more uncertain than those of N70.

\subsection{Spatial Distribution}

The H$\alpha$ image of N185 shows a complex structure.  While its outer rim in the northern hemisphere and the southwest quadrant form a coherent shell morphology, its southeast quadrant appears to show a distinct, smaller shell morphology and additional blowout features extending outward to at least 2$'$ (corresponding to 30 pc) from the shell rim of N185.  \ion{H}{1} mapping in the vicinity of N185 indicates a lack of associated \ion{H}{1} gas \citep{2002AJ....123..255O}.  N185 thus seems to be located within the cavity of the supergiant shell LMC-7 \citep{1980MNRAS.192..365M}.

The X-ray images of N185 are produced with similar methods as we do for N70. The useful observations of N185 were cut short by high background flares; thus, the images of N185 have much lower S/N than those of N70. The diffuse X-ray emission from N185 (Figure\ 2) is distributed mostly within the shell interior, but does extend to the southeast exterior.  There is a particularly good correlation with the features in the southeast quadrant: the diffuse X-ray emission peaks within the interior shell and extends into the southeast blowout structure even beyond the H$\alpha$ filaments. In the band of 0.4--0.7 keV, the diffuse emission in the blowout region rivals that at the peak within the shell.

\begin{figure}
\includegraphics[width=0.3\textwidth,angle=270,clip=]{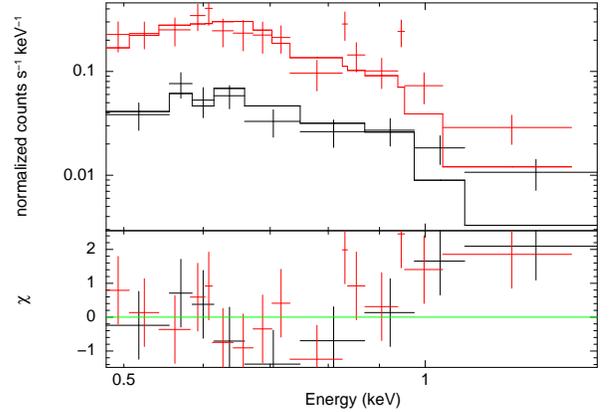}
\caption{\emph{XMM-Newton} EPIC X-ray spectra of entire N185. The merged MOS1 and MOS2 spectra and PN spectra are plotted in black and red, respectively. The best-fit model is overplotted.}
\label{figure8} 
\end{figure}

Comparisons between the X-ray images of N185 and the \ion{H}{1} column density map or channel maps do not show any \ion{H}{1} enhancements in X-ray-faint regions within N185. Thus, the distribution of the diffuse X-ray emission reflects the distribution of the hot gas instead of the foreground absorption.  The hot gas in N185 exists mostly in the small shell in the southeast quadrant and extends into the blowout region.  Comparisons between the X-ray and H$\alpha$ images of N185 show anticorrelation between H$\alpha$ and X-ray surface brightnesses in two regions.  First, the bright east--west oriented H$\alpha$ filament that bisects N185 corresponds to a narrow strip of low surface brightness in the band of 0.7--1.0 keV (in Figure \ref{figure2}(c)).  Second, the area of low X-ray surface brightness between the blowout region and the shell interior emission region corresponds to the H$\alpha$ shell rim (in Figure \ref{figure2}(b)).  If these anticorrelations are caused by absorption of X-rays by the ionized gas shell of N185, the blowout must be on the far side of the shell and the east--west H$\alpha$ filament must be caused by a density enhancement on the near side of the shell.

\subsection{Spectral Analysis}  

The X-ray surface brightness of N185 is low.  While Figure \ref{figure2} shows apparent spatial variations in spectral properties of the diffuse X-ray emission in N185, there are not enough counts to extract spectra separately from the shell interior and the blowout.  We thus analyze the spectrum of the diffuse X-ray emission for the entire N185, but determine the hardness ratios for the blowout and shell interior separately and compare them.  

Because of the small number of counts, it is difficult to make satisfactory model fits to the X-ray spectra of N185. After experimenting with different models, we find the hot gas in N185 to be best described by an absorbed apec model of optical thin hot gas in collisional ionization equilibrium without metal overabundance (see also Table 1 for the fitting results).  To make crude estimates of the plasma properties, we have adopted the hydrogen column density derived from the $E(B-V)$ = 0.14 of the O9 star in N185 \citep{1996ApJ...465..231O} as the absorption column density, $N_{\rm H}\sim 2.2\times 10^{21}$ H-atom cm$^{-2}$.  With this fixed absorption column, the best-fit model gives a gas temperature of $kT \sim 0.2$ keV and a normalization factor {\it A} of $\sim 1.4 \times 10^{-3}$ cm$^{-5}$. Assuming the hot gas originates from the interior of the shell structure with a 50 pc radius and a He/H number ratio of 10\%, the hydrogen density is $n_{\rm H}$ = 0.05 $f^{-1/2}$ cm$^{-3}$, and the hot gas mass is  $\sim$ $5.9\times10^{2}$ $f^{1/2}$ $M_\odot$. The filling factor $f$ is most likely greater than 0.1, and the hot gas mass is greater than 100 $M_\odot$. This large mass is consistent with the hot gas being shock-heated interstellar gas without metal overabundance.  Although line features cannot be discerned in Figure 8, a non-thermal ({\it powerlaw}) model yields an uncomfortably high photon index ($\sim$4.8) and large $\chi^2_\nu$ ($\sim$1.7).

The blowout region alone has too few counts for spectral analysis (17, 29, and 80 counts in MOS1, MOS2, and PN, respectively).  It is only possible to compare the spectral hardness of the blowout with that of the shell interior.  Using the MOS1+MOS2+PN counts, we find the (0.7--1.0 keV)/(0.4--0.7 keV) count ratio is 0.24$\pm$0.07 for the blowout and 0.66$\pm$0.09 for the shell interior.  It thus appears that the blowout region has softer X-ray emission than the shell interior.  Softer X-ray emission is associated with cooler plasma temperature, which is expected for the hot gas expanding into regions with lower pressure.

\section{Discussion}

\subsection{Diffuse X-Ray Emission Expected from Superbubbles}

In a classical wind-blown bubble or a superbubble, shock-heated gas exists and emits in X-rays \citep{1977ApJ...218..377W,1995ApJ...450..157C}; however, the X-ray emission from a superbubble is faint, unless off-center SNRs impact the dense superbubble shell walls \citep{1990ApJ...365..510C}. Both N70 and N185 encompass multiple OB stars, and thus their shells are naturally candidates of superbubbles.  Whether the shell is recently heated by an off-center SN can be determined by comparing the observed diffuse X-ray emission with that expected from a quiescent superbubble.

The X-ray luminosity of a quiescent, i.e., X-ray-dim, superbubble can be formulated as
$$L_{\rm x}=(8.2\times 10^{27} ~{\rm erg}~{\rm s}^{-1}) \xi I(\tau) n_0^{10/7} R_{\rm pc}^{17/7} V_{5}^{16/7}~,$$
where $V_5$ is the superbubble expansion velocity in units of km s$^{-1}$, $R_{\rm pc}$ is the superbubble radius in parsecs, $n_0$ is the hydrogen number density in the ambient interstellar medium in units of H-atom cm$^{-3}$, $\xi$ is the metal abundance relative to the solar value, and $I(\tau)$ is a dimensionless integral with a value of $\sim$2 \citep{1995ApJ...450..157C}.

The LMC metal abundance is roughly 1/3 solar, and thus $\xi \sim$ 1/3.  The densities of the ionized gas shells of N70 and N185 have both been estimated to be $\sim$6.5 H-atom cm$^{-3}$ \citep{1996ApJ...467..666O}.  Assuming that the shell consists of swept-up ambient medium, then 
$$n_{0} \simeq 3(\frac{\bigtriangleup R}{R_{\rm s}})n_{\rm s}~,$$
 
\begin{figure}
\centering
\includegraphics[width=0.45\textwidth,angle=0,clip=]{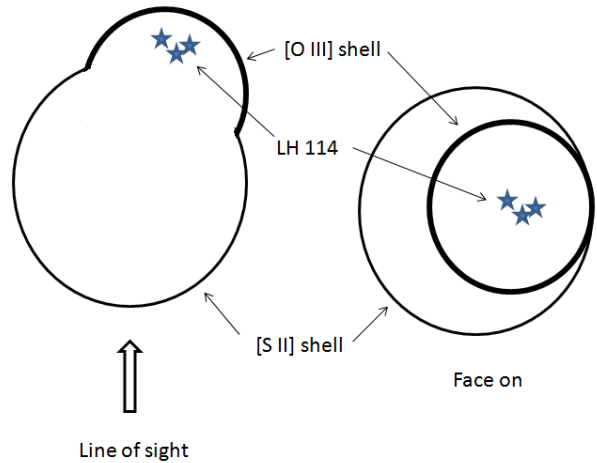}
\caption{Illustrations of two views of the three-dimensional structure 
of N70. The sketch on the left shows a side view and the right shows a face-on view. The thick line delineates the portion of the shell running into a denser interstellar medium.}
\label{figure9} 
\end{figure}
 
where $R_{\rm s}$ is the shell radius, $\bigtriangleup R$ is the shell thickness, and $n_{\rm s}$ is the shell density.  Both N70 and N185 have shell radii of $\sim$50 pc and fractional shell thicknesses of 0.06$\pm$0.01; thus, the ambient density would be $n_0 \sim$ 1.2$\pm$0.2 H-atom cm$^{-3}$.  We adopt the shell expansion velocities determined from our echelle observations: $\le35$ km s$^{-1}$ for N70 and $\ge$70 km s$^{-1}$ for N185.  Using these shell parameters and ambient density, we expect the X-ray luminosity of a quiescent superbubble to be $\le(3.2\pm0.7)\times10^{35}$ erg s$^{-1}$ for N70 and $\ge (1.6\pm0.4)\times10^{36}$ erg s$^{-1}$ for N185.  The luminosity of diffuse X-ray emission from N70, determined from our observation and model fits to its spectrum, is $\sim 6.1\times10^{35}$ erg s$^{-1}$, higher than that expected from a quiescent superbubble, which can still be fainter if the expansion velocity of N70 is much lower than 35 km s$^{-1}$. In contrast, the observationally determined diffuse X-ray luminosity of N185 ($\sim 1.9\times10^{35}$ erg s$^{-1}$) is an order of magnitude lower than that expected from a quiescent superbubble. 


\subsection{Physical Nature of N70}

N70 encompasses the OB association LH114, and thus the UV flux and fast stellar winds from the massive stars are expected to photoionize the ambient medium and dynamically sweep the medium into a superbubble.  Its superbubble structure is complicated by the existence of a molecular cloud that is coincident with the bright patches of H$\alpha$  emission and the OB association LH114 \citep{2008ApJS..178...56F}.  The lack of pronounced absorption of the diffuse X-ray emission indicates that the dense molecular cloud is on the far side of the superbubble. Massive young stellar objects have been found in N70, and are associated with the molecular cloud \citep{2009ApJS..184..172G}.  It appears that the star formation in N70 is still ongoing.

We suggest that the OB association LH114 is formed in the molecular cloud but near its surface; thus, N70 expands into the dense cloud on one side and forms a blister on the tenuous side, as shown by the illustrations in Figure \ref{figure9}.  The observed H$\alpha$ morphology can be reproduced by observing with the line-of-sight nearly perpendicular to the cloud surface.  Given this configuration, the [\ion{O}{3}]-bright shell structure corresponds to the dense hemisphere expanding into the dense cloud; the blister side of the superbubble has a higher [\ion{S}{2}]/H$\alpha$ ratio with its west rim projected near the west rim of the [\ion{O}{3}] shell rim and its east rim further east from the [\ion{O}{3}] rim.

The brightest optical emission from N70 corresponds to the photoionized dense hemisphere, whose free-free emission dominates in the radio wavelengths, resulting in a thermal spectral index.  The X-ray luminosity from N70, higher than that expected from a quiescent superbubble, suggests a recent occurrence of a SN that heats the hot gas in the superbubble interior.  The age of the SN is not known, but must be younger than the age of the superbubble.  A recent numerical three-dimensional hydrodynamic simulation of N70 has also concluded that the morphology, dynamics, and X-ray emission of N70 are best explained by a recent SN in a superbubble \citep{2011ApJ...733...34R}.

\subsection{Physical Nature of N185}

N185 has a less dense environment than N70.  It is not associated with a molecular cloud or ongoing star formation, and it has no known OB association.  If N185 is a superbubble, its X-ray luminosity is an order of magnitude lower than expected.  While the X-ray emission from wind-blown bubbles has been observed to be lower than expected from bubble models \citep[e.g.,][]{1996ApJ...467..666O,2002AJ....124.3325N,2001ApJ...553L..69C,2003ApJ...599.1189C}, it should be noted that the superbubble nature of N185 has not been rigorously established.  The stellar population in N185 has been studied by \citet{1996ApJ...465..231O}.  It is shown that the currently most massive stars in N185 include 10 early-B stars  (12--15 $M_\odot$) and 4 late-O stars (20--25 $M_\odot$), which are not powerful sources of fast stellar winds for superbubble formation.  Extrapolating the observed mass function to higher masses, it is expected that two stars in the mass range of 40--60 $M_\odot$ have exploded as SNe \citep{1996ApJ...467..666O}.  The total mechanical energy injected by these two massive stars, both stellar winds and SN explosions, would by far exceed that from the combined wind energy provided by the lower-mass stars in N185.


The expansion velocity of N185 is high and irregular.  The largest expansion velocity registered on the receding side of the shell reaches $\sim$200 km s$^{-1}$.   Such a high expansion velocity is not seen in any superbubble in the LMC; however, such fast expansion velocity and the irregular expansion are common characteristics of SNRs in the LMC \citep{1988AJ.....95.1111C}. Furthermore, the radio emission from N185 is nonthermal,  and the X-ray luminosity of N185 is consistent with that of an evolved SNR in a low-density medium, such as SNR 0450-70.9 \citep{2004ApJ...613..948W}.  Therefore, we suggest that a SNR produced the diffuse X-ray emission, the smaller shell-like structure in the southeast quadrant, and the blowout structure of N185.  The large shell (100 pc in diameter) is likely to be formed by an earlier event, possibly related to the first SN in N185.  The occurrence of two SNe is consistent with the expectation from the observed stellar population in N185, and they would indeed have played a dominant role in the formation of N185's shell structure.

\section{Summary}

N70 and N185 are two large [\ion{S}{2}]-bright shells in the LMC.  We have obtained \emph{XMM-Newton} X-ray observations and high-dispersion long-slit echelle spectroscopic observations of these two shells.  These \emph{XMM-Newton} observations, much more sensitive than the previous \emph{Einstein} or \emph{ROSAT} observations, allow useful spatial and spectral analyses for the first time.  Diffuse X-ray emission is detected from both shells and their apparent blowout regions, and the X-ray spectra are soft with plasma temperatures of $kT \sim 0.20\pm0.02$ keV. 

The X-ray luminosity of N70 is higher than that expected from a quiescent superbubble model; thus, the interior of N70 is most likely heated by a recent SN.  The X-ray luminosity of N185 is an order of magnitude lower than that expected from a quiescent superbubble with the size and expansion velocity of N185. The hot gas in N185 is most likely originating from a SNR, as it shows nonthermal radio emission, fast and irregular expansion, and its X-ray luminosity is consistent with that of an evolved SNR.  The large shell of N185 may be associated with an earlier SN event.  These explanations are supported by the observed stellar population in these two large shell nebulae.

\begin{acknowledgments}

This work has partially benefited from the support from NSFC grants 11233001 and 11203013, the 973 Program grant 2009CB824800, SRFDP of China 20120091110048 and 20110091120001 by the Educational Ministry of China, and grants by the 985 Project of NJU and the Advanced Discipline Construction Project of Jiangsu Province.  R.M.W. and Y.H.C. acknowledge the support of NASA grant NNX06AH65G.  We also thank Wei Sun and Ping Zhou for helpful discussions.

\end{acknowledgments}




\end{document}